\documentclass[aps,showpacs,twocolumn, superscriptaddress,prb]{revtex4-1}

\usepackage{graphicx}
\usepackage{amsmath,amssymb,amsfonts}
\usepackage{textcomp}
\usepackage{hyperref}
\usepackage{gensymb}
\usepackage{verbatim}
\usepackage{amsmath}
\hypersetup{breaklinks=true,colorlinks=true,urlcolor=black}

\begin{document}
	
	\title{Superconducting properties of  Rh$_{9}$In$_4$S$_4$ single crystals} 
	
	\author{Udhara S. Kaluarachchi}  
	\affiliation{Ames Laboratory, U.S. DOE, Iowa State University, Ames, Iowa 50011, USA}
	\affiliation{Department of Physics and Astronomy, Iowa State University, Ames, Iowa 50011, USA}
	\author{Qisheng Lin}  
	\affiliation{Ames Laboratory, U.S. DOE, Iowa State University, Ames, Iowa 50011, USA}
	\author{Weiwei Xie$^{\dagger}$}
	\affiliation{Ames Laboratory, U.S. DOE, Iowa State University, Ames, Iowa 50011, USA}
	\affiliation{Department of Chemistry, Iowa State University, Ames, Iowa 50011, USA}
	\author{Valentin Taufour}
	\affiliation{Ames Laboratory, U.S. DOE, Iowa State University, Ames, Iowa 50011, USA}
	\author{Sergey L. Bud'ko}  
	\affiliation{Ames Laboratory, U.S. DOE, Iowa State University, Ames, Iowa 50011, USA}
	\affiliation{Department of Physics and Astronomy, Iowa State University, Ames, Iowa 50011, USA}
	\author{Gordon J. Miller}
	\affiliation{Ames Laboratory, U.S. DOE, Iowa State University, Ames, Iowa 50011, USA}
	\affiliation{Department of Chemistry, Iowa State University, Ames, Iowa 50011, USA}
	\author{Paul C. Canfield}
	\affiliation{Ames Laboratory, U.S. DOE, Iowa State University, Ames, Iowa 50011, USA}
	\affiliation{Department of Physics and Astronomy, Iowa State University, Ames, Iowa 50011, USA}

	\begin{abstract}
	The synthesis and crystallographic, thermodynamic and transport properties of single crystalline Rh$_{9}$In$_4$S$_4$ were studied. The resistivity, magnetization and specific heat measurements all clearly indicate bulk superconductivity with a critical temperature, $T_{c}\sim$\,2.25\,K. The Sommerfeld coefficient $\gamma$\ and the Debye temperature($\varTheta_{\textrm{D}}$) were found to be 34\,mJ\,mol$^{-1}$ K$^{-2}$  and 217\,K respectively. The observed specific heat jump, $\Delta C/\gamma T _{c}$\,=\,1.66, is larger than the expected BCS weak coupling value of 1.43. Ginzburg-Landau (GL) ratio of the low temperature GL-penetration depth, $\lambda_\text{GL}$$\approx$\,5750\AA, to the GL-coherence length, $\xi_\text{GL}$$\approx$\,94\AA, is large: $\kappa$\,$\sim$\,60. Furthermore, we observed a peak effect in the resistivity measurement as a function of both temperature and magnetic field. 

	\end{abstract}
\maketitle

	\section{Introduction}
	Transition-metal chalcogenides show diverse physical states such as charge density wave\cite{Wilson1974,Harper1975,Monceau1976PRL,DiSalvo1976}, superconductivity\cite{Morosan2006Nature,Nagata1992,Naren2008,Hsu2008,Sakamoto2006}, ferromagnetism\cite{Natarajan1988,Carteaux1991} and semi-conducting behavior\cite{Grant1975,KoujiTaniguchi2012}. The ability to change their properties by doping\cite{Acrivos1971,Morosan2006Nature,Goto1997} or pressure\cite{Mizuguchi2008,Yomo2005PRB} has recently attracted great attention. Of specific interest, some members of metal-rich chalcogenides\cite{Michener1943,Peacock1950,Brower1974,Natarajan1988},  $A_{2}M_{3}X_{2}$ ($A$=Sn,Pb,In,Tl and Bi; $M$=Co,Ni,Rh and Pd; $X$=S and Se) are superconducting at low temperatures\cite{Sakamoto2006,Sakamoto2008,Lin2012}. Interestingly Bi$_{2} $Rh$ _{3}$Se$ _{2}$\cite{Sakamoto2007} is a superconductor that shows a possible high-temperature ($\sim$\,240\,K) charge density wave transition. In contrast, the isostructural  Bi$_{2} $Rh$ _{3}$S$ _{2}$\cite{Natarajan1988,Kaluarachchi2015PRB} has a high temperature structural phase transition, but remains non-superconducting down to 0.5\,K, and the neighboring Bi$_{2} $Rh$ _{3.5}$S$ _{2}$\cite{Kaluarachchi2015PRB} has no structural phase transition, but becomes superconducting at $T_{c}\approx$\,1.7\,K. The discovery of superconductivity in Bi$_{2} $Rh$ _{3.5}$S$ _{2}$ motivated us to extend our exploration for superconducting compounds to the Rh-In-S system, which has not yet been fully investigated with only one compound, Rh$_{3}$In$_2$S$_2$\cite{Natarajan1988}, reported.  

	In this article, we present details of the crystal growth and characterization of the transition-metal chalcogenide superconductor Rh$_{9}$In$_4$S$_4$. Measurements of transport properties, magnetization and specific heat confirm bulk superconductivity of Rh$_{9}$In$_4$S$_4$ at $T_{c}\sim$\,2.25\,K and we report other superconducting properties from the above measurements. The upper critical field, $\mu_0H_\text{c2}$, shows good agreement with the Helfand-Werthamer (HW) theory. We also present the observation of a peak effect in this material by means of transport measurements.

	\section{Experimental Methods}
	
		Single crystals of  Rh$_{9}$In$_4$S$_4$ were produced using a solution growth technique\cite{CANFIELD1992,Canfield2001,Lin2012}. A mixture of elemental Rh, In and S was placed in a 2 mL fritted alumina crucible\cite{Petrovic2012,Canfield2016} with a molar ratio of Rh:In:S\,=\,55:22.5:22.5 and sealed in a silica ampule under a partial pressure of high purity argon gas. The sealed ampule was heated to 1150\,\celsius\, over 12 hours and held there for 3 hours. After that, it was cooled to 950\celsius\, over 50 hours and excess liquid was decanted using a centrifuge. Single crystals of Rh$_{9}$In$_4$S$_4$ grew as tetragonal rods with typical size of $\sim0.5\times0.5\times2$\,mm$^3$ as shown in the inset of Fig.\,\ref{XRD}(b).

		Single crystal X-ray diffraction data were collected using a Bruker SMART APEX II CCD area-detector diffractometer\cite{SMART2003} equipped with Mo  K$_{\alpha}$ ($\lambda$ = 0.71073\AA) radiation. Integration of intensity data was performed by the SAINT-Plus program,  absorption corrections\cite{Blessing1995} by SADABS , and least-squares refinements by SHELXL\cite{Sheldrick2002}, all in the SMART software package. Lattice parameters were refined using single crystal diffraction data and are summarized in Table\,\ref{Tb_lattice}. Atomic coordinates and displacement parameters with full site occupancy for Rh$_9$In$_4$S$_4$ are derived from the single crystal diffraction and given in Table\,\ref{Tb_Wyckoff}. Powder X-ray diffraction data were collected using a Rigaku Miniflex II diffractometer at room temperature (Cu K$_{\alpha}$ radiation). Samples for powder X-ray diffraction was prepared by grinding single crystals and spreading them onto a thin grease layer coated single crystal Si, zero background puck.  Powder X-ray diffraction data were analyzed using the GSAS\cite{Toby2001,GSAS} program.
		
		\begin{table}[htb!]
			\caption{Lattice parameters of Rh$_{9}$In$_4$S$_4$ at 293\,K. All values are from single crystal diffraction data. }
			\begin{tabular}{|p{3cm}|p{3cm}|}
				\hline
				Formula      & Rh$_{9}$In$_4$S$_4$ (293\,K)     \\ 
				\hline
				Formula weight &  1485.89            \\
				$ Z $-formula units &  2 \\
				Space group   & I4/m m m (139)     \\ 
				$a$ ($\textrm{\AA}$) &  7.7953(3)    \\
				$c$ ($\textrm{\AA}$) &  8.8583(3)         \\
				Volume ($\textrm{\AA}^{3}$) &  538.25(5)   \\
				Density (g/cm$^{3}$) &  9.339    \\
				
				\hline
				
			\end{tabular}
			\label{Tb_lattice}
		\end{table}
		
		\begin{table}[htb!]
			\caption{Atomic coordinates and equivalent isotropic displacement parameters of Rh$_{9}$In$_4$S$_4$ at 293\,K.}
			\begin{tabular}{|p{.75cm} | p{.8cm}| p{1.1cm} | p{1.3cm}| p{1.3cm} | p{1.3cm}| p{1.1cm}|}
				\hline
				\hline
				\centering Atom &	\centering Wyck &	\centering Symm. &	\centering x &	\centering y & \centering	z & $U _{eq} $ ($\textrm{\AA}^{2}$) \\
				\hline
				In1 & 4e & 4mm   & 0  & 0      	 & 0.1872(3) & 0.027(1)  \\
				In2 & 4d & -4m2   & 0.5  & 0      	 & 0.25 & 0.019(1) \\
				Rh1 & 8f & ..2/m & 0.25    & 0.25 & 0.25       & 0.067(1) \\
				Rh2 & 8i & m2m   & 0.3011(3)  & 0     	 & 0 & 0.019(1) \\
				Rh3 & 2b & 4/mmm   & 0          & 0  & 0.5       & 0.012(1) \\
				S   & 8h & m.2m   & 0.2079(7)  & 0.2079(7)  & 0.5  & 0.038(2)  \\ \hline	 
				\hline				
			\end{tabular}
			\label{Tb_Wyckoff}
		\end{table}	
	
		The ac resistivity ($f$\,=\,17\,Hz) was measured as a function of temperature and field by the standard four probe method in a Quantum Design (QD), Physical Property Measurement System (PPMS) instrument. Four Pt wires with  diameters of 25\,$\mu$m were attached to the samples using Epotek-H20E silver epoxy or DuPont 4929N silver paint. The contact resistance was $\sim0.5\Omega$. The specific heat was measured by using the relaxation method in the PPMS. The $^{3}$He option was used to obtain measurements down to 0.4 K. The total uncertainty of the specific heat data is $\sim$\,$5\%$.  The DC magnetization measurements were performed in a QD, Magnetic Property Measurement System (MPMS).

	\section{Results}
	
		\subsection{Structure}
		
		\begin{figure}[htb!]
			\begin{center}
				\includegraphics[width=8.5cm]{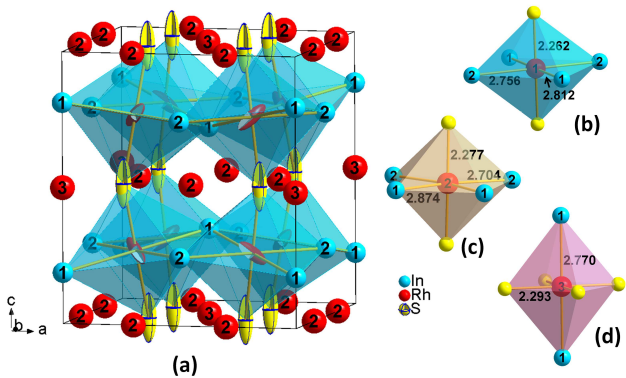}
			\end{center}
			\caption{\label{srtuc_Ref} (a) Unit cell of Rh$_{9}$In$_4$S$_4$, with Rh1-centered octahedra shaded in blue. Rh1 and S atoms are shown as ellipsoids with 90$\%$ probability. (b)-(d) show the detailed configurations of Rh1-, Rh2-, and Rh3-centered octahedra, together with representative bond distances. Numbers overlaid with colored spheres denoted atoms listed in Table.\,\ref{Tb_Wyckoff}.} 
		\end{figure}
		
		\begin{figure}[h!]
			\centering
			\includegraphics[width=8cm]{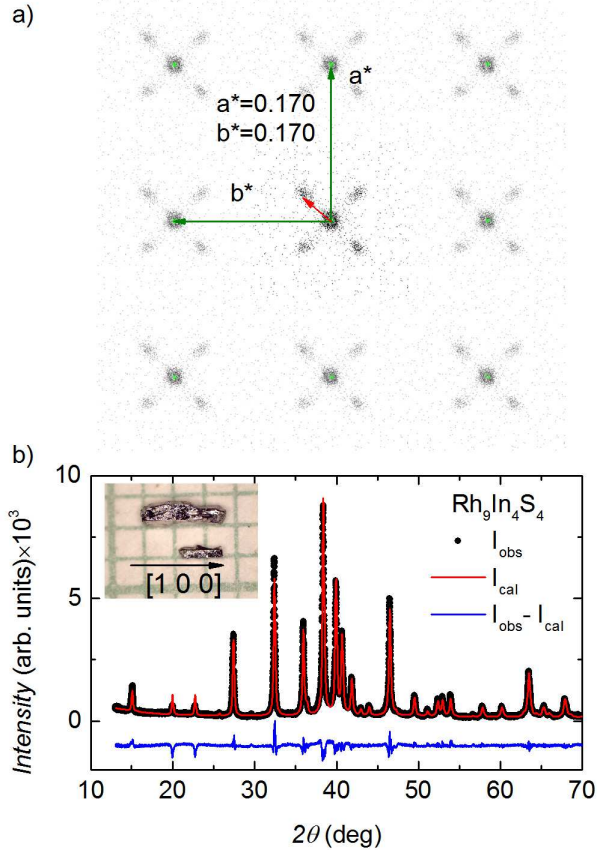}
			\caption{(a) Reciprocal lattice viewed along (001) zone for  Rh$_{9}$In$_4$S$_4$. Green dots denote a 2\,$\times$\,2 main lattice reflections of the average structure given in Table.\,\ref{Tb_lattice}; four-fold, more diffuse, clusters around the green dots denote weak reflections of modulated lattice, with a modulation vector of [0.17 0.17 0] (red arrow). (b) Powder diffraction pattern of  Rh$_{9}$In$_4$S$_4$. The red line represents the calculated diffraction pattern based on lattice parameters obtained from the single crystal diffraction analysis. The blue line represents the difference between the experimental and calculated intensities. The inset shows a photo of single-crystalline Rh$_{9}$In$_4$S$_4$  on a millimeter grid and the arrow indicates the [1 0 0] direction.} 
			\label{XRD}
		\end{figure}

		Figure\,\ref{srtuc_Ref}(a) shows the unit cell of the average structure of Rh$_{9}$In$_4$S$_4$, which is described by the tetragonal space group I4/mmm. The  powder X-ray diffraction pattern of a ground, phase pure, single crystal of Rh$_{9}$In$_4$S$_4$ is shown in Fig.\,\ref{XRD}\,(b). According to single crystal X-ray diffraction analyses (Table\,\ref{Tb_lattice}), Rh$_{9}$In$_4$S$_4$ crystallizes in the tetragonal symmetry I4/mmm ($a$\,=\,7.7953(3)\,\AA, $c$\,=\,8.8583(3)\,\AA,). Powder pattern was fitted with LeBail refinement and obtained $R_p$\,=\,7.3$\%$ as shown in Fig.\,\ref{XRD}\,(b). Lattice parameters obtained from this measurement are in good agreement (less than 0.2$\%$) with single crystals data. 
		
		The composition was refined as Rh$_{9}$In$_4$S$_4$, consistent with Rh$_{2.2(1)}$InS measured by electron probe micro-analyzer (EPMA). The lattice parameters and atomic coordinates are listed in Table.\,\ref{Tb_lattice} and Table.\,\ref{Tb_Wyckoff}, respectively. As in the case of Bi$_2$Rh$_3$S$_2$\cite{Kaluarachchi2015PRB}, all Rh atoms in Rh$_{9}$In$_4$S$_4$  are  six-coordinated  forming slightly distorted octahedra. Of these, Rh1 and Rh2 are surrounded by 4 In and 2 S atoms, whereas Rh3 by 2 In and 4 S atoms, see Fig.\,\ref{srtuc_Ref}\,(b)-(d). Whereas Rh2 and Rh3 octahedra sit on edges or face centers of the unit cell, Rh1 octahedra are located at (1/4 1/4 1/4) and equivalent sites. The arrangement of these Rh1-centered octahedra is similar to those in Perovskites, except that they share edges (In1-In2) in the $ab$ plane and vertices along $c$, whereas in Perovskites only corners are shared. Apparently, the pivot-and-rock motion of these Rh1-centered octahedra is restricted in $ab$ plane, resulting in ellipsoid elongation for the Rh1 and S atoms, Fig.\,\ref{srtuc_Ref}\,(a). Viewing along the $c$ axis, Rh1 and S atoms form a zig-zag chains extending along $c$, and their disorder can easily lead to structure modulation, as observed in many other structures, e.g., Sc$_4$Mg$_x$Cu$_{15-x}$Ga$_{7.5}$\cite{QishengLin2008}

		Indeed, careful examination of reciprocal space from single crystal intensity data confirms that the structure is a modulated structure. With a cut-off intensity of $3\sigma$, we were able to identify a modulation vector of [0.17 0.17 0], see Fig.\,\ref{XRD}\,(a). However, due to the weakness of the modulation reflections, no model of any modulated structure has been acceptable (so far). Since the results of an average structure refinement are sufficient for our current discussions, herein we will no longer focus on the detailed modulated structure, but rather the averaged structure.


		
		\subsection{Physical properties of Rh$_{9}$In$_4$S$_4$}

\begin{figure}[htb!]
		\centering
		\includegraphics[width=8.5cm]{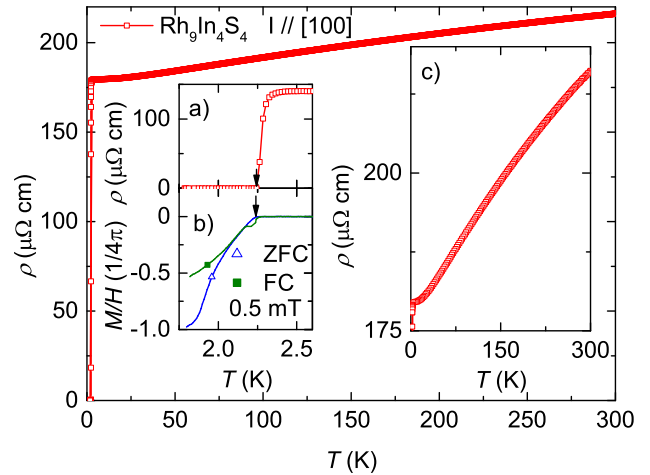}
		\caption{ Temperature dependent resistivity of Rh$_{9}$In$_4$S$_4$ along $\textrm{[1~0~0]}$.  The insets show (a) the typical  superconducting transition feature in resistivity data and the arrow indicates the offset criteria, which was used to obtained the $T_\textrm{c}$\,$\approx$\,2.25\,K; (b) shows the ZFC and FC $M/H$ of Rh$_{9}$In$_4$S$_4$ and the arrow represents the onset criteria which was used to obtain $T_\textrm{c}$\,$\approx$\,2.24\,K; (c) shows the normal state resistivity in expanded scale and weak negative curvature visible at high temperatures.} 
	\label{Rho}
	\end{figure}

	Figure\,\ref{Rho} shows the temperature dependent resistivity of Rh$_{9}$In$_4$S$_4$ for current flowing along the $\textrm{[1~0~0]}$ direction. The resistivity decreases monotonically with decreasing temperature, showing metallic behavior and a clear sharp transition to zero resistivity below $T_\text{c}$\,=\,2.25\,K, indicating a superconducting transition of this material (Fig.\,\ref{Rho}\,(a)). The residual resistivity ratio (RRR) ($\rho_{\textrm{300\,K}}$/$\rho_{\textrm{5.5\,K}}$)  is 1.2. Figure\,\ref{Rho}\,(b) show the zero-field-cooled (ZFC) and field-cooled (FC) magnetization data of Rh$_{9}$In$_4$S$_4$ and clearly indicates over 95\,{\%} shielding fraction at 1.8\,K. Also in the experimental data we can see a weak negative curvature in resistivity at higher temperatures as shown in Fig.\,\ref{Rho}\,(c).

\begin{figure}[htb!]
		\centering
		\includegraphics[width=8.5cm]{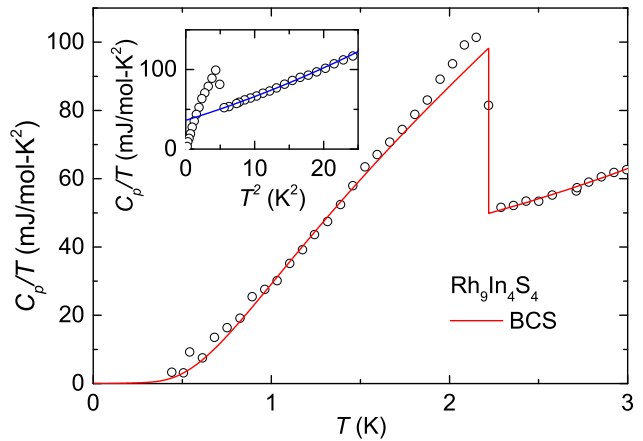}
		\caption{ Low temperature $C_{p}/T$ vs $T$ of Rh$_{9}$In$_4$S$_4$.  Red solid line represents the BCS calculation. The inset shows the $C_{p}/T$ vs $T^{2}$ graph which  was used to obtain $\gamma$, $\beta$ and $\delta$ values.  The blue solid line in the inset represents a fit with  $C_{p}/T$\,=\,$\gamma+\beta T^{2}$.} 
		\label{CP}
	\end{figure}

	Figure\,\ref{CP} shows the low temperature specific heat data of Rh$_{9}$In$_4$S$_4$. A fit for $C_{p}/T$\,=\,$\gamma+\beta T^{2}$ from 2.3 to 3.5\,K for the normal state, as shown in the inset of Fig.\,\ref{CP}, yielded the Sommerfeld coefficient, $\gamma$\,=\,34\,mJ\,mol$^{-1}$\,K$^{-2}$ (or $\sim$\,2\,mJ\,$\textrm{mol-atomic}^{-1}$\,K$^{-2}$), $\beta$\,=\,3.22\,mJ\,mol$^{-1}$\,K$^{-4}$.  From the $\beta$ value we estimate the Debye temperature, $\varTheta_{\textrm{D}}$\,=\,217\,K, using the relation $\varTheta_{\textrm{D}}$\,=\,$(12\pi^{2}n\textrm{R}/(5\beta))^{1/3}$, where $n$ is the number of atoms per formula unit and $R$ is the universal gas constant.

	With a large superconducting volume fraction, the specific heat data are expected to reveal a clear anomaly at $T_{c}$. We obtained the $T_{c}$\,$\approx$\,2.22\,K  and the specific heat jump of $\Delta C$\,=\,125\,mJ\,mol$^{-1}$\,K$^{-1}$ by using an equal entropy construction to the low temperature specific heat data. Given that  $T_{c}$\,$\approx$\,2.25\,K from the resistivity, $T_{c}$\,$\approx$\,2.24\,K from the magnetization and $T_{c}$\,$\approx$\,2.22\,K from the specific heat, we can state  $T_{c}\sim$\,2.25\,K for Rh$_{9}$In$_4$S$_4$.
	
	$\Delta C/\gamma T _{c}$\,=\,1.66 is an important measure of electron-phonon coupling strength, which is stronger here, than the BCS weak-coupling limit of 1.43. The red colored line if Fig.\,\ref{CP} represents the BCS\cite{Bardeen1957,Tinkham1996} calculation for the weak limit.

	The electron-phonon coupling constant $\lambda_{e-ph}$ can be estimated from the McMillan equation\,\cite{McMillan1968} for the superconducting transition temperature, for phonon mediated superconductors,
	
	\begin{equation}
		T_{c}= \frac{\varTheta_{\textrm{D}}}{1.45} \textrm{exp}  \left[  -\frac{1.04(1+\lambda_{\textrm{e-ph}})}{\lambda_{\textrm{e-ph}}-\mu^{*}(1+0.62\lambda_{\textrm{e-ph}}) }\right]       
		\label{Eq_McMillan}
	\end{equation}
	
	\noindent where $\mu^{*}$, the Coulomb pseudopotential, has a  value often between 0.1 and 0.2 and usually is taken as 0.13\,\cite{McMillan1968}. Similar values of $\mu^{*}$ have been used in other Rh-containing chalcogenides\cite{Sakamoto2008,Sakamoto2007,Naren2008}. Using $\varTheta_{\textrm{D}}$\,=\,217\,K and $T_{c}$\,=\,2.22\,K we estimated  $\lambda_{\textrm{e-ph}}$\,=\,0.56. A difference of $\mu$ from the assumed  value of 0.13 will give a different value of  $\lambda_{\textrm{e-ph}}$. For example,  $\lambda_{\textrm{e-ph}}$\,=\,0.5 if $\mu$\,=\,0.1 and  $\lambda_{\textrm{e-ph}}$\,=\,0.71 if $\mu$\,=\,0.2. The value of  $\lambda_{\textrm{e-ph}}$ indicates the sample is an intermediate coupled superconductor\cite{McMillan1968}. The ratio between BCS-coherence length and mean free path  can be written as\cite{Kittel2005} $\xi_{BCS}/l$=$(0.18\hbar n\rho_0 e^2)/(k_{\text{B}}T_cm^*)$. Using the values of $\rho_0$\,=\,180\,$\mu\Omega$\,cm, $T_\text{c}$=2.24\,K,  $m^*$\,=\,$m_e(1+\lambda_{\textrm{e-ph}})$ and assuming electron density for typical metal, n\,$\approx$\,$10^{27}-10^{28}$\,m$^{-3}$, we can calculate $\xi_\text{BCS}/l$\,$\approx$\,20-200. This value is much greater than one, indicating that  Rh$_{9}$In$_4$S$_4$  is unambiguously in the dirty limit.
	 
	\begin{figure}[htb!]
		\centering
		\includegraphics[width=8.5cm]{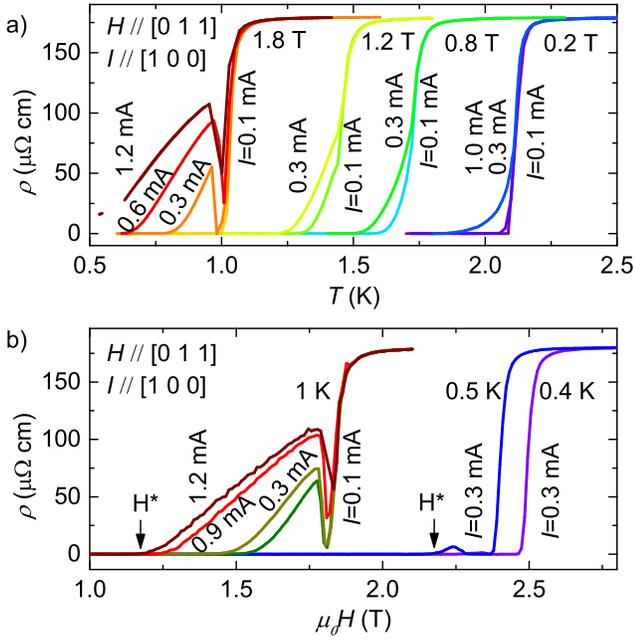}
		\caption{Low temperature resistivity as a function of (a) temperature and (b) field for several applied currents for  $H \parallel [0~1~1]$ configuration where current flow along the [1\,0\,0] direction.} 
		\label{current_depn}
	\end{figure}

	\begin{figure}[th!]
		\centering
		\includegraphics[width=8.5cm]{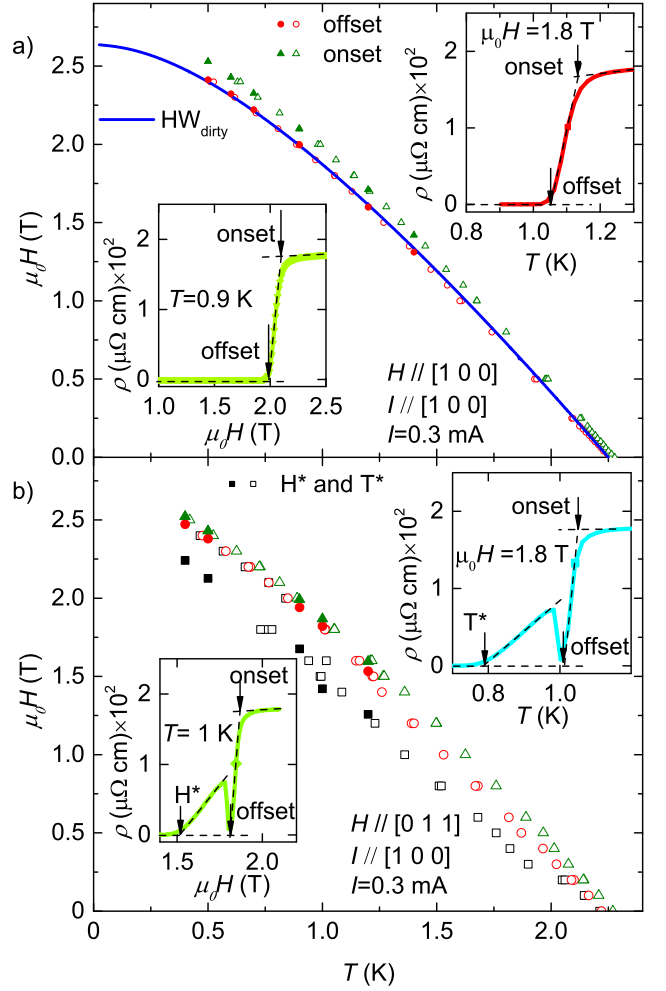}
		\caption{Upper critical field $H_\text{c2}$ vs $T$ of Rh$_{9}$In$_4$S$_4$ for (a) $H \parallel I$ and (b) $H \perp I$ configurations with current along $a$-axis. Lower and upper insets show the criteria which was used to obtained the data points. Solid and open symbols represent the data obtained by field scans and temperature scans respectively. The blue solid line in (a) indicates the HW calculations for the dirty limit. In (b), lower and upper insets show the resistivity anomalies due to the peak effect and $H^*$ and $T^*$ data represent by solid and open black squares.} 
		\label{HC2}
	\end{figure}

	In some type-II superconductors, a sharp maximum in the temperature or field dependence of critical current observed below $H_\text{c2}(T)$ is called  the "peak effect"\cite{Berlincourt1959PRev,Berlincourt1961PRL,Berlincourt1963PRev,DeSorbo1964PRev,Pippard1969}. Several mechanisms have been proposed for the explanation of the peak effect such as matching mechanism\cite{Raffy1972}, elementary pinning by weakly superconducting regions\cite{Campbell1972,Livingston1966}, reduction in elastic moduli of the flux line\cite{Larkin1979} and the synchronization of the flux line lattice\cite{Pippard1969,Kramer1973}. However, the underlying physics is not yet fully understood so far. Figure\,\ref{current_depn}\,(a) shows the temperature dependence of resistivity data for several applied fields and measuring currents for an $H \parallel [0~1~1]$ configuration. For $I$\,=\,$0.1$\,mA, at field of $\sim$\,0.2\,T the transition is quite sharp. However for moderate fields, such as 0.8 or 1.2\,T, the resistivity data show a kink in between onset and offset of the transition and, for further increase of the field (e.g. 1.8\,T), the resistivity drops to zero before increasing to a finite value. For higher currents these anomalies in resistivity become more prominent and at higher fields, resistivity just show a dip before increasing and then gradually going to zero. The anomalies in the resistance are seen more clearly in the resistivity vs. magnetic field at fixed temperatures for the different values of measuring current ($H \parallel [0~1~1]$ configuration) as shown in Fig.\,\ref{current_depn}\,(b). With the increase of field, the resistivity is essentially zero up to a critical field ($H^*$) and it starts to increase. For further increase of field, the resistivity reaches maximum and goes to zero again before increasing to the normal state. For larger current, $H^*$ decreases and height of the peak increases. It should be  emphasized that there were no anomalies in the transition observed when the current is parallel to the magnetic field. Similar peak effect data has been observed in several compounds including Nb\cite{Autler1962PRL}, CeRu$_2$\cite{Dilley1996,Sato1996,Hedo1998}, NbSe$_2$\cite{Higgins1996}, V$_3$Si\cite{Chaudhary2001}, ReCr\cite{Stognei1997}, Yb$_3$Rh$_4$Sn$_{13}$\cite{Tomy1997,Tomy1997PRB} and MgB$_2$\cite{Lyard2002PRB,Welp2003PRB,Lee2008} . However we observed this effect at low current densities, $j$, between 0.3-3\,A\,cm$^{-2}$ in contrast with other superconducting materials ($j\gg$10\,A\,cm$^{-2}$)\cite{Dilley1996,Welp2003PRB,Hedo1998,Lee2008,Sato1996,Stognei1997,Tomy1997}.

	The $H-T$ phase diagrams obtained from the low temperature $R$ vs. $T$ and $R$ vs. $H$ measurements are presented in Fig.\,\ref{HC2} for $I$\,=0.3\,mA. Figure\,\ref{HC2}\,(a) and (b) show $H \parallel [1~0~0]$ and $H \parallel [0~1~1]$ configurations and the upper and lower insets of each figure show the criteria used to determine $H_\text{c2}$ and $T_c$. Open and closed symbols represent data obtained from temperature scans and field scans respectively. As noted above, the peak effect is only detectable in the $H \parallel [0~1~1]$ configuration. Comparison of $H_\text{c2}$ in $H \parallel [1~0~0]$ and $H \parallel [0~1~1]$ configurations indicates virtually isotropic $H_\text{c2}$ behavior.

	For a one band, type II superconductor the orbital upper critical field is given by Helfand-Werthamer (HW)\cite{Helfand1966} theory and can be estimated from $H_\text{c2}$\,=\,-A$T_\textrm{c}$d$H_\text{c2}$/d$T$, where A\,=\,0.73 for clean limit and 0.69 for the dirty limit. The slope of the curve in the vicinity of $T_c$ (for the offset criteria) is -1.69\,T\,K$^{-1}$. Using this value, the calculated $\mu_{0}H_\text{c2}$ is 2.78\,T for the clean limit and 2.62\,T for the dirty limit. The blue solid line in Fig.\,\ref{HC2}\,(a) represents the calculated HW curve for the dirty limit (Ref.\,\onlinecite{Helfand1966} equation 24), and it shows a good agreement with the experimental data.

	 By using both normal state and superconducting state specific heat data, one can obtain the thermodynamic critical field, $\mu_{0}H_{c}$($T$) as a function of temperature from equation\,\ref{Eq_ExpHC}

	\begin{equation}
	\frac{\mu_{0} V_{m} H_\text{c}(T)^{2}}{2}=\int\limits_{T}^{T_\text{c}}\Delta S(T')dT'       
	\label{Eq_ExpHC}
	\end{equation}

	\noindent in which $\Delta S(T)$ is the entropy difference between the normal and superconducting states and $V_{m}$ (16.2$\times$10$^{-5}$ m$^{3}$ mol$^{-1}$) is the molar volume. The calculated value of $\mu_{0}H_{c}$($T$\,=\,0) is 25\,mT for Rh$_{9}$In$_4$S$_4$ and it is much smaller than $\mu_{0}H_\text{c2}(0)$=2.62\,T as expected for type II superconductor. The value of $\mu_{0}H_\text{c2}(0)$
	is well below the  Pauli paramagnetic limit\cite{Clogston1962,Chandrasekhar1962} of $\mu_{0}H_{\textrm{c2}}^{p}(0)$\,=\,$1.84\,T_{\textrm{c}}$\,=\,4.1\,T, suggesting  an orbital pair-breaking mechanism.  
	
	We also can estimate the Ginzburg-Landau(GL) coherence length\cite{Tinkham1996}, $\xi_\text{GL}$\,$\approx$\,94\,$\textrm{\AA}$ by using the relation $\text{d}(\mu_{0}H_\text{c2}(T_c))/\text{d}T$\,=\,-$\Phi_{0}/({2\pi\xi_\text{GL} ^{2}T_c)}$, in which $\Phi_{0}$ is the quantum flux and estimated  $\text{d}(\mu_{0}H_\text{c2}(T_c))/\text{d}T$ to be -1.69\,T\,K$^{-1}$ near $T_c$. London penetration depth for the dirty limit can be written as\cite{Parks1969},
	\begin{equation}
	\lambda^{-2}(T)\,=\,\frac{4\pi^{2}\Delta(T)}{(c^{2}\hbar\rho_{0,cgs})}tanh(\Delta(T)/(2T))  
	\label{eq_1}
	\end{equation}
	\noindent where $\Delta(T)$ is temperature dependence of the superconducting gap energy and all parameters are in cgs units.
	
	\noindent Near $T_\text{c}$,
	\begin{eqnarray}
		\Delta^2(T\rightarrow T_\text{c})&=\frac{8\pi^2T_c^2k_B^2}{7\zeta(3)}(1-\frac{T}{T_c}) \nonumber\\
		\lambda^{-2}(T\rightarrow T_\text{c})\,&=\,\frac{4\pi^{2}}{c^{2}\hbar\rho_{0,cgs}}\frac    {\Delta^2(T\rightarrow T_\text{c})}{2T_ck_B}  \nonumber\\
		\lambda^2(T\rightarrow T_\text{c})&=\frac{\lambda^2_{GL}}{(1-\frac{T}{T_c})} 
		\label{eq_nearTc}
	\end{eqnarray}
	\noindent where $\zeta$ is the Riemann zeta function and $\zeta(3)$\,$\approx$\,$1.202$. The expression in Eq.\,\ref{eq_nearTc} can be reduced and converted to SI units as $\lambda(T\!\!\!\rightarrow\!\! T_c)$\,=\,$0.00064\sqrt{\rho_{0}/(T_c(1-T/T_c))}$\,=\,$\lambda_\text{GL}/\sqrt{(1-T/T_c)}$, where $\rho_0$ is in SI units. Using $\rho_0$\,=\,$180\times10^{-8}\,\Omega\,m$ and $T_\text{c}$\,=\,$2.25\,K$, we can obtained $\lambda_\text{GL}$\,=\,$0.00064\sqrt{\rho_0/T_\text{c}}\approx5750\text{\AA}$. Based on $\xi_\text{GL}$ and $\lambda_\text{GL}$ values, estimated GL parameter $\kappa$\,=\,$\lambda_\text{GL}$/$\xi_\text{GL}$ is $\sim$\,61. Jump of the specific heat and slope of $H_{c2}$ at $T\textrm{c}$ can be calculated from the Rutger's relation\cite{Rutgers1934,Welp1989PRL}
	\begin{equation}
	\Delta C/T_\text{c}=(1/8\pi\kappa^{2})(\textrm{d}H_\text{c2}/\textrm{d}T)^{2}|_{T_\text{c}}
	\label{eq_Rutgers}	
	\end{equation} 
	\noindent where $\Delta C$  in units of erg\,cm$^{-3}$\,K$^{-1}$ and slope  $H_{c2}$ in units of Oe\,K$^{-1}$. Using the molar volume, $V_m$\,=\,162.1\,cm$^3$\,mol$^{-1}$ we obtain the converted $\Delta C$\,=\,7710\,erg\,cm$^{-3}$\,K$^{-1}$. From the value of $H_{c2}$ slope near $T_\text{c}$, d$H_{c2}$/d$T$\,=\,16900\,Oe\,K$^{-1}$, we obtained a similarly large $\kappa$ value of 57. This considerably large $\kappa$ value indicates that Rh$_{9}$In$_4$S$_4$ is an extreme type II superconductor.  A summary of the measured and derived superconducting state parameters for Rh$_{9}$In$_4$S$_4$ is given in Table.\ref{Tb_summary}

	\begin{table}[th!]
	\caption{Measured and derived superconducting and relevant normal-state parameters for Rh$_{9}$In$_4$S$_4$}
	\begin{tabular}{|p{5cm}|c|}
			\hline
			\multicolumn{1}{|c|}{Rh$_{9}$In$_4$S$_4$ property}                    & ~~~Value~~~ \\ \hline
			$T_\text{c}$ (K)                                                      & 2.25(2)                       \\
			$\gamma$ (mJ\,mol$^{-1}$\,K$^{-2}$)  								  & 34.0(4)                         \\
			$\beta$ (mJ\,mol$^{-1}$\,K$^{-4}$)    								  & 3.22(5)                       \\
			$\Theta_D$ (K)                                                        & 217                        \\
			$\Delta C$ (mJ\,mol$^{-1}$\,K$^{-1}$) 								  & 125(4)                        \\
			$\Delta C/\gamma T _{c}$                                              & 1.66(6)                       \\
			$\lambda_{e-ph}$                                                      & 0.56                       \\
			$\rho_0$ ($\mu\Omega$\,cm)                                            & 180                        \\
			$H_{c2}(T=0)$ (T) (clean limit)                                       & 2.78                       \\
			$H_{c2}(T=0)$ (T) (dirty limit)                                       & 2.62                       \\
			$H_{c2}^p(T=0)$ (T)        		                                      & 4.1                       \\
			$H_{c}(T=0)$ (mT)                                                     & 25                         \\
			$\xi_\text{BCS}/l$                                                    & 20-200 					   \\
			$\xi_\text{GL}$ ($\AA$)                                               & 94   					   \\
			$\lambda_{GL}$ ($\AA$)                                                & 5750   					   \\
			$\kappa=\lambda_{GL}/\xi_\text{GL}$                                   & 61    					   \\
			$\kappa$ (from Rutger's relation)                                     & 57    					   \\
			\hline
			\hline
	\end{tabular}
		\label{Tb_summary}
	\end{table}
\newpage	 

		\section{Conclusions}
		We report the synthesis, crystal structure and characterization (such as resistivity, magnetization and specific heat) of superconducting Rh$_{9}$In$_4$S$_4$ with a bulk superconducting transition of $T_\text{c}\sim$\,2.25\,K and large value of GL parameter $\kappa\sim$\,60. Rh$_{9}$In$_4$S$_4$ is found to be a type II and intermediate-coupling superconductor. The calculated values for the Sommerfeld coefficient and the Debye temperature are 34\,mJ\,mol$^{-1}$\,K$^{-2}$ and 217\,K respectively. The temperature dependence of the specific heat shows a larger jump  $\Delta C/\gamma T _\text{c}$\,=\,1.66 at $T_\text{c}$, than the BCS weak coupling limit. The upper field critical  shows a good agreement with the HW theory. We have shown a direct evidence of a peak effect from resistivity in  Rh$_{9}$In$_4$S$_4$ single crystals as a function of both temperature and magnetic field. Further studies of superconducting properties and peak effect of this material will be useful to understand the nature of their mixed state and pinning properties.

		\section*{ACKNOWLEDGMENTS}
		We would like to thank W.~Straszheim, Z.~Lin and K.~Sun for experimental assistance and A.~Kreyssig, R.~Prozorov, V.\,G.~Kogan, T.~Kong, M.~Kramer and R.\,J.~Cava (Princeton University) for useful discussions. This work was supported by the U.S. Department of Energy (DOE), Office of Science, Basic Energy Sciences, Materials Science and Engineering Division. The research was performed at the Ames Laboratory, which is operated for the U.S. DOE by Iowa State University under contract No. DE-AC02-07CH11358. V.T. is partially supported by Critical Material Institute, an Energy Innovation Hub funded by U.S. DOE, Office of Energy Efficiency and Renewal Energy, Advanced Manufacturing Office.

	$\dagger$ Current address: Department of Chemistry, Princeton University, Princeton, NJ 08544, USA.


\newpage	
\pagebreak


%

\end{document}